\documentclass[12pt]{article}

\usepackage{amsmath}
\usepackage{amssymb}
\usepackage{graphicx}
\usepackage{xcolor}
\usepackage{array}
\usepackage{url}
\usepackage{hyperref}
\usepackage{natbib}

\usepackage[strict]{changepage} 
\usepackage{framed} 
\usepackage{romannum}
\usepackage{enumitem}
\definecolor{grey}{gray}{0.95}
\newenvironment{formal}{%
  \MakeFramed{\advance\hsize-\width\FrameRestore}%
  \noindent\hspace{-4.55pt}
  \begin{adjustwidth}{}{7pt}%
  \vspace{2pt}\vspace{2pt}%
}
{%
  \vspace{2pt}\end{adjustwidth}\endMakeFramed%
}

\newcommand{\blind}{0}

\begin{document}


\pagenumbering{arabic}

\def\spacingset#1{\renewcommand{\baselinestretch}%
{#1}\small\normalsize} \spacingset{1}


\if0\blind
{
  \title{\bf Designing a Data Science simulation \\ with $\mathcal{MERITS}$: A Primer}
  \author{
    \vspace{-1.2cm} \\
    Corrine F. Elliott\textsuperscript{a}\thanks{The authors gratefully acknowledge partial support from NSF grant DMS-2413265; NSF Grant 2023505 for the Foundations of Data Science Institute (FODSI); and NSF grant MC2378 to the Institute for Artificial CyberThreat Intelligence and OperatioN (ACTION).  CFE acknowledges NSF Research Training Grant 1745640; TMT acknowledges the NSF Graduate Research Fellowship Program (DGE-2146752); and MB acknowledges Deutsche Forschungsgemeinschaft (DFG, project \#509149993, TRR 374).  CFE thanks Giles Hooker for multiple reviews and D. Stephen Voss for a thorough edit of an early draft and critical discussion.  Any errors belong to the authors alone.}
    , \\
    James P.C. Duncan\textsuperscript{b}\thanks{Authors contributed equally to this manuscript.}
    , Tiffany M. Tang\textsuperscript{c}\textsuperscript{\dag}, \\
    Merle Behr\textsuperscript{d}\thanks{Authors contributed equally to this manuscript.}
    , Karl Kumbier\textsuperscript{e}\textsuperscript{\ddag}, \\ 
    and Bin Yu\textsuperscript{a,b} \\
    \vspace{-0.25cm}\scriptsize\textsuperscript{a} Department of Statistics, University of California, Berkeley, CA, USA; \\
    \vspace{-0.25cm}\scriptsize\textsuperscript{b} Graduate Group in Biostatistics, University of California, Berkeley, CA, USA; \\
    \vspace{-0.25cm}\scriptsize\textsuperscript{c} Department of Statistics, University of Michigan, MI, USA; \\
    \vspace{-0.25cm}\scriptsize\textsuperscript{d} Faculty of Informatics and Data Science, University of Regensburg, Regensburg, Germany \\
    \scriptsize\textsuperscript{e} Department of Pharmaceutical Chemistry, University of California, San Francisco, CA, USA
  }
  \maketitle
} \fi

\if1\blind
{
  \bigskip
  \bigskip
  \bigskip
  \begin{center}
    \title{\bf Designing a Data Science simulation \\ with $\mathcal{MERITS}$: A Primer}
  \end{center}
  \medskip
} \fi

\vspace{-1cm}
\begin{abstract}
Simulations play a crucial role in the modern scientific process.  Yet despite (or due to) this ubiquity, the Data Science community shares neither a comprehensive definition for a ``high-quality'' study nor a consolidated guide to designing one.  Inspired by the Predictability-Computability-Stability (PCS) framework for `veridical' Data Science, we propose six $\mathcal{MERITS}$ that a simulation study should satisfy. ($\mathcal{M}$\textit{odularity} and $\mathcal{E}$\textit{fficiency} support the computability of a study, encouraging clean and flexible implementation. $\mathcal{R}$\textit{ealism} and $\mathcal{S}$\textit{tability} address the conceptualization of the research problem: How well does a study predict reality, such that its conclusions generalize to new data/contexts?  Finally, $\mathcal{I}$\textit{ntuitiveness} and $\mathcal{T}$\textit{ransparency} encourage good communication and trustworthiness of study design and results.)  Drawing an analogy between simulation and cooking, we moreover offer (a) a conceptual framework for thinking about the anatomy of a simulation `recipe'; (b) a baker's dozen in guidelines to aid the Data Science practitioner in designing one; and (c) a case study demonstrating the practical utility of our framework by using it to autopsy a preexisting simulation study.  With this ``PCS primer'' for high-quality Data Science simulation, we seek to distill and enrich the best practices of simulation across disciplines into a cohesive recipe for trustworthy, veridical Data Science.  

\end{abstract}

\noindent%
\vfill

\newpage
\spacingset{1.5} 

\tableofcontents

\newpage
\section{Introduction: Why simulate in Data Science?} \label{sec:intro}

Simulations play a crucial role in modern Data Science research.  Quality studies facilitate the scientific process by defining a realistic training ground where researchers can (a) sound out hypotheses to distinguish those worth pursuing; (b) stress-test fledgling theories; and (c) model natural systems under specific or rare conditions.  Complementary to other avenues of scientific exploration (\textit{e.g.}, empirical investigation, inductive reasoning), simulations thus offer three main advantages: a streamlined simulacrum of reality, unfettered control of experimental conditions, and the ability to explore obscure empirical scenarios.

\medskip Simulations allow researchers to interrogate scientific hypotheses faster and less discriminately than by logical or empirical validation.  Such efficiencies particularly benefit methodological stress-testing (\textit{e.g.}, \citealt{Suzuki, Maas, Strasburg, Chen, Yu-Yao}) and clinical-trial design (\textit{e.g.}, \citealt{Holford00, Holford10, Bonate, Girard, Orloff, Meurer}).  In each context, we can simulate various realistic settings and observe the results directly -- for multiple scenarios in parallel -- to learn faster than by inductive or mathematical reasoning alone, and across settings where an analytical approach might be intractable or require strong assumptions.  This combined breadth and efficiency allows us to explore various scenarios and develop a nuanced understanding of the study system in reasonable time.

\medskip Beyond mere speed, simulation offers precise control over experimental conditions.  Suppose we wish to compare two (or more) analytical techniques (as in \citealt{Suzuki, Chen, Yu-Yao, Haller, Kolgatin}).  Choosing a superior method requires that neither receives an undue advantage, and simulation offers a fair arena for competition: identical tests in identical computing environments using identical or analogous data.  Such control over conditions \textit{in silico} also allows us to emulate circumstances that are difficult to observe empirically.  This feature benefits the study of models for rare events such as genetic mutations or natural disasters~\citep{Hartmann, ElGheriani, Webber}.  Data describing such systems are inherently scant, such that we risk over-fitting if we build models using (one type of) real-world data alone.  We can mitigate this risk by augmenting our data with realistically generated synthetic data.

\medskip Data scientists across many disciplines try to exploit computation for rapid scientific progress.  But speed must be tempered by careful design to promote grounded, reliable conclusions.  Despite (or due to) the modern ubiquity of simulations, the collective Data Science community shares neither agreed criteria for a ``high-quality'' study nor guidance for designing one, giving rise to a crisis of reproducibility \citep{Luijken}.  Neophyte data scientists learn design practices \textit{ad hoc} from hearsay or by trial-and-error.  Such scattershot learning is fallible (because one might not stumble into good practice without concerted guidance) and inefficient (because one learns fitfully).  We seek to raise the baseline for simulation quality, and facilitate meeting that standard, by distilling and enriching the best practices of simulation across disciplines into a recipe for trustworthy Data Science.

\medskip A review of related work (Section~\ref{sec:disc}) reveals no unified paradigm for conceptualizing, designing, and appraising simulations.  Our work builds most directly from the Predictability-Computability-Stability (PCS) framework for veridical (loosely, ‘truthful’) Data Science, which encourages strong connections between simulation and reality (conceptualization) through the eponymous three pillars~\citep{Yu-Kumbier}. Here \textit{predictability} invokes the use of prediction in validating models against real-world expectations; \textit{computability} concerns the logistics and fidelity of translating a real-world problem into a computational one; and \textit{stability} encompasses not only statistical variation but also perturbations in human decision-making and analytical pipelines. These principles are undoubtedly valid concerns, but the framework lacks tangible guidance for designing a satisfactory study. Other groups advocate narrowly for one or two specific quality criteria [appraisal], such as algorithmic accountability, interpretability, or reproducibility~\citep{Rule, Krafczyk}.  Existing design guidelines vary widely in breadth and scale, even within a manuscript. These contributions aggregate to hint at a unifying simulation paradigm but do not articulate it, and none of the closest analogues has won widespread acceptance in the field.

\medskip To fill this niche, we offer a ``PCS primer'' for high-quality Data Science simulation.  We illustrate our design philosophy using analogies to a widely relatable theme: the preparation and consumption of food. Herein we describe the anatomy of a simulation plan, akin to the four components of a recipe (Section~\ref{sec:components}); propose six criteria for high-quality simulation (Section~\ref{sec:merits}); and provide a checklist of guidelines for designing one (Section~\ref{sec:primer}). Section~\ref{sec:casestudy} consolidates these elements by dissecting a preexisting simulation study.\footnote{See Appendices~\ref{sec:LSSFind}-\ref{sec:VRL} for additional examples.}  We conclude by connecting our primer to competing and complementary antecedents (Section~\ref{sec:disc}).

\section{Four components of a simulation recipe} \label{sec:components}

\begin{figure}[ht]
    \centering
    \includegraphics[width=0.8\linewidth]{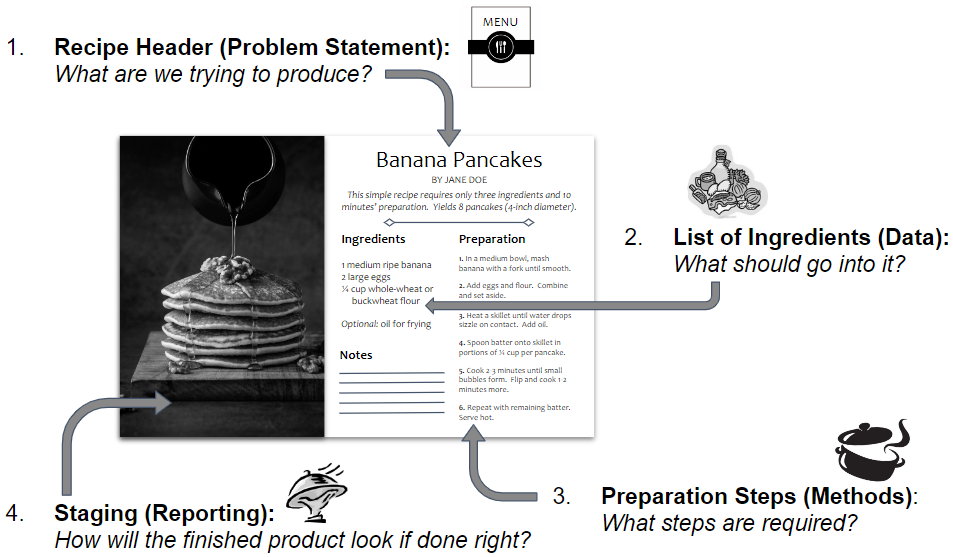}
    \caption{Components of a (simulation) recipe.}
    \label{fig:recipe}
\end{figure}

A simulation plan comprises four broad components: the problem statement, data, methods, and reporting (Figure~\ref{fig:recipe}).  For best transparency and rigor, these components should be planned and documented upfront; in practice, they might evolve with new information, but such evolution likewise should be recorded.  A well-documented study can be replicated by another scientist to arrive at analogous results.  In these ways, as in the parts themselves (below), the motivation and anatomy of study documentation resemble those of an idealized cooking recipe.  Just as a detailed recipe captures the full plan for preparing a dish, so should the project documentation comprise the full plan for a simulation study:

\begin{enumerate} 
\item[A.] \textbf{Recipe Header (Problem Statement)}: \textit{What are we trying to produce?}  \\ Our problem statement resembles a \textit{recipe header}, which gives an overview and description of the dish.  It specifies our motivation and trajectory. (Section~\ref{sec:header})
\item[B.] \textbf{List of Ingredients (Data)}: \textit{What should go into it?} \\ Data (and/or data-generating processes) comprise the \textit{ingredients} of our simulation study.  These are the elements on which we act, whose quality and appropriateness for the intended dish influence the caliber of our final product. (Section~\ref{sec:ingredients})
\item[C.] \textbf{Preparation Steps (Methods)}: \textit{What steps are required?} \\ Study methods constitute our \textit{preparation techniques} for the intended dish.  Much as a recipe dictates both equipment and cooking specifications (\textit{e.g.}, techniques, times, temperatures), so does our study methodology encompass models, computational parameters, and evaluation metrics alike.  Our choices must suit the ingredients, intended product, and prior knowledge of the (culinary or scientific) field.  (Section~\ref{sec:prep})
\item[D.] \textbf{Staging (Reporting)}: \textit{How will the finished product look if done right?} \\ Communicating results is analogous to \textit{staging} a dish.  Presentations may comprehend visual, qualitative, and/or quantitative summaries, depending on our audience, but should enhance rather than detract from the experience of consuming our product.  Reporting also includes consolidating materials for reproducing the work, akin to the chef annotating a recipe to make it easier or better for next time. (Section~\ref{sec:staging})
\end{enumerate}

\noindent Although the purpose and components of each plan are similar between the culinary and data sciences, an important distinction exists: Home cooks often work from existing recipes.  By contrast, the first step in conducting a simulation study often is to \textit{write our own recipe}.  To that end, Section~\ref{sec:merits} proposes six desiderata to strive for in designing a simulation, and Section~\ref{sec:primer} offers a checklist of guidelines for succeeding.

\section{Flavor of a high-quality simulation: Six \texorpdfstring{$\mathcal{MERITS}$}{MERITS}} \label{sec:merits}

The term ``simulation study'' applies to many and diverse computational experiments, but these endeavors often share common design elements and may be judged by similar quality standards.  We move that a study should exhibit the following minimal $\mathcal{MERITS}$, detailed in Table~\ref{tab:merits}: $\mathcal{M}$odularity, $\mathcal{E}$fficiency, $\mathcal{R}$ealism, $\mathcal{I}$ntuitiveness, $\mathcal{T}$ransparency, and $\mathcal{S}$tability.

\newcolumntype{L}[1]{>{\raggedright\let\newline\\\arraybackslash}p{#1}}
\begin{table}[htbp]
\begin{center}
\resizebox{\columnwidth}{!}{
\begin{tabular}{||l | m{0.75\textwidth} | b{0.18\textwidth}||} 
 \hline
 \textbf{Merit} & \textbf{Definition} & \textbf{Guidelines} \\ [0.5ex] 
 \hline\hline
 $\mathcal{M}$odular & \textbf{Written in self-contained and logically partitioned code segments to ease interpretation, auditing, and adaptation to new settings.}  Modules (\textit{e.g.}, data generation, model fitting) are exchangeable with minimal revision to the analysis pipeline.  & \ref{g1}, \ref{g8} \\ 
 \hline
 $\mathcal{E}$fficient & \textbf{Streamlined computationally and conceptually.} Simulations are scalable, minimizing run-times over sufficient replicates without loss of functionality.  Designs are judicious in varying parameters\footnote{To afford a comprehensive view of the problem space, without ``boiling the ocean.''} and accounting for pre-existing work.\footnote{Incorporating benchmarks for comparing across studies, without burdensome redundancies.} & \ref{g1}, \ref{g4}, \ref{g9} \\
 \hline
 $\mathcal{R}$ealistic & 
 \textbf{Faithful to the physical world, as best understood from established theory and real-world data.} Simplifications required for study comprehension and tractability do not undermine the relevance of simulation results to reality.\footnote{In particular, designs accommodate realistic sources of randomness (\textit{e.g.}, sampling, subject heterogeneity, missing data, measurement error) and other practical limitations inherent to the data/model.} & \ref{g1}, \ref{g3}, \ref{g4}, \ref{g6}, \ref{g10} \\
 \hline
 $\mathcal{I}$ntuitive & \textbf{Sensible to the intended audience and, in a general sense, to a reasonably comprehensive readership.}\footnote{This standard applies to all research products (\textit{e.g.}, simulation architecture, documentation, analysis).}  Although models seeking to capture real-world complexities might be abstruse, a resourceful layperson can identify the study assumptions, parse its logical flow, and gauge the soundness of its conclusions. & \ref{g1}, \ref{g2}, \ref{g5}, \ref{g9}, \ref{g11} \\
 \hline
 $\mathcal{T}$ransparent & \textbf{Documented thoroughly and candidly.} Records provide all information needed to conduct or assess the simulation study, including the rationale for any human judgement calls, study strengths and limitations, and key takeaways. & \ref{g1}, \ref{g2}, \ref{g3}, \ref{g7}, \ref{g8}, \ref{g11}, \ref{g12}, \ref{g13} \\
 \hline
 $\mathcal{S}$table & \textbf{Reproducible/replicable}\footnote{Ideally the data scientist publishes materials to re-execute (reproduce) a study, but at least documents all details pertinent to re-creating (replicating) the work~\citep{Barba, NASEM}.} \textbf{and externally valid.} Conclusions are reasonably consistent across executions, random seeds, users, computers, and sensible perturbations to the analysis pipeline (\textit{e.g.}, human decisions in data cleaning). & \ref{g1}, \ref{g5}, \ref{g6}, \ref{g7}, \ref{g10}, \ref{g12}, \ref{g13} \\ [1ex] 
 \hline
 \multicolumn{3}{L{1.2\textwidth}}{\small $^2$To afford a comprehensive view of the problem space, without ``boiling the ocean.''} \\
 \multicolumn{3}{L{1.2\textwidth}}{\small $^3$By incorporating benchmarks enough for across-study comparisons, but without burdensome redundancies.} \\
 \multicolumn{3}{L{1.2\textwidth}}{\small $^4$In particular, designs incorporate realistic sources of randomness (\textit{e.g.}, sampling, subject heterogeneity, missing data, measurement error) and other practical limitations inherent to the data/model.} \\
 \multicolumn{3}{L{1.2\textwidth}}{\small $^5$This standard applies to all research products (\textit{e.g.}, simulation architecture, documentation, analysis).} \\
 \multicolumn{3}{L{1.2\textwidth}}{\small $^6$Ideally the data scientist publishes materials to re-execute (reproduce) a study, but, at minimum, documents all details pertinent to re-creating (replicating) the work~\citep{Barba, NASEM}.} \\
\end{tabular}
}
\end{center}
\caption{\texorpdfstring{$\mathcal{MERITS}$}{MERITS}: Six desiderata for a simulation study of high quality.}\label{tab:merits}
\end{table}

\medskip These $\mathcal{MERITS}$ resonate strongly with the PCS framework that helped inspire them~\citep{Yu-Kumbier}.  $\mathcal{M}$odularity and $\mathcal{E}$fficiency support the \textit{computability} of a study, encouraging clean and flexible implementation of simulations designed for tractability and scalability.  $\mathcal{R}$ealism and $\mathcal{S}$tability encourage faithful translation between real-world research problems and the mathematical constructs employed \textit{in silico}: How well does a simulation \textit{predict} reality, such that its conclusions generalize to new data/contexts?  Results that are \textit{stable} across perturbations to the data, models, and human decisions are more likely to coincide with reality.  Finally, we encourage $\mathcal{I}$ntuitiveness and $\mathcal{T}$ransparency (auxiliary to the PCS acronym, but recommended implicitly under the ``PCS documentation'' standard) for clear and trustworthy communication of study designs and results.

\medskip Although often unacknowledged, these principles quietly underpin many high-quality simulation studies.  Yet even unassailable virtues can be difficult to uphold without tangible guidance for achieving them.  We therefore follow with a ``baker's dozen'' in specific guidelines (Section~\ref{sec:primer}) for raising the baseline quality of Data Science simulation studies. 

\section{Guidelines for designing a study with \texorpdfstring{$\mathcal{MERITS}$}{MERITS}} \label{sec:primer}

This section presents thirteen guidelines for cooking up a simulation study with $\mathcal{MERITS}$: one overarching precept and three guidelines to help shape each recipe component.  These recommendations are not meant to be memorized -- although obeying them might, with practice, become second nature -- but rather to serve as a reference for the conscientious practitioner in shoring up his or her simulation recipe prior to implementation.  All six $\mathcal{MERITS}$ should pervade the entire design process, but for conciseness, we limit discussion to the two desiderata most pertinent to each guideline.  We conclude each subsection with a pertinent example from the literature wherein these guidelines were upheld.

\vspace{-0.3cm}
\begin{center}
\rule{300pt}{0.5pt}
\end{center}
\vspace{-0.5cm}

\begin{formal}
\begin{enumerate}[series=guidelines,label=\roman*.,ref=\roman*]
    \item Plan and document in advance all feasible simulation specifications and contingencies (\textit{e.g.}, architecture, data, mathematical machinery, sources of randomness).\label{g1}
\end{enumerate}
\end{formal}

\noindent This first precept most nearly summarizes our simulation design philosophy.  Common wisdom holds that the key to quick-and-easy recipe execution is to plan ahead and perform as much preparatory work as possible in advance of the cooking itself.  We call this process the \textit{mise en place} (roughly, ``putting in position'').  It entails reading the recipe thoroughly; assembling the necessary equipment; and priming the ingredients (rinsing, defrosting, chopping, measuring).  It means having all elements ``cooking-ready'' to reduce the number of moving parts and risk of critical mistakes in the final execution. If the procedure includes a tricky step, we might plan a contingency in case our first attempt should fail.  The effort we invest in planning and preparing, we recoup from the seamless, less-stressful cooking process.  We also reduce the probability of having nothing to eat in the end.

\medskip This philosophy extends naturally to the simulation context, in which we anticipate and prepare for all facets of our study.  We first plan our study `recipe,' documenting not just the obvious statistical particulars (data-generating processes, models, and evaluation metrics) but also computational specifications (code architecture and libraries, number of replications, random seeds or procedure for choosing them, parallelization strategies, data storage and access, server connectivity, etc.), and ensuring that all are ``simulation-ready'' prior to implementation.  We moreover should try to anticipate hiccups and plan contingencies: For example, if a model fails to converge, do we record a failure or re-fit the model with a different algorithm?  By anticipating these particulars, we minimize \textit{ad hoc} decisions during implementation and, by extension, the risk of bias or error.  Such deliberate forethought also helps in designing a study with all of our desired $\mathcal{MERITS}$.

\begin{table}[htbp]
\begin{center}
\resizebox{0.9\columnwidth}{!}{%
\begin{tabular}{||l | m{0.75\textwidth} | l ||} 
 \hline
 \multicolumn{3}{||l||}{\textbf{Overall}} \\ 
 \hline\hline
 \ref{g1} & Plan and document in advance all feasible simulation specifications and contingencies.
 & $\mathcal{MERITS}$ \\
 \hline\hline
 \multicolumn{3}{||l||}{\textbf{Recipe Header (Problem Statement)}} \\ 
 \hline
 \ref{g2} & Describe as concretely as possible the research question(s) your study seeks to answer. 
 & $\mathcal{T}$ransparent, $\mathcal{I}$ntuitive \\
 \hline
 \ref{g3} & Connect your study to reality, while acknowledging its limitations honestly. 
 & $\mathcal{R}$ealistic, $\mathcal{T}$ransparent \\
 \hline
 \ref{g4} & Recognize and accommodate the wider research context for your study (\textit{e.g.}, domain knowledge or body of literature). 
 & $\mathcal{E}$fficient, $\mathcal{R}$ealistic \\
 \hline\hline
 \multicolumn{3}{||l||}{\textbf{List of Ingredients (Data)}} \\ 
 \hline
 \ref{g5} & Test methods in a (data) context as reminiscent as possible of their intended deployment. 
 & $\mathcal{I}$ntuitive, $\mathcal{S}$table \\
 \hline
 \ref{g6} & Incorporate real-world data and/or accepted scientific theory in generating synthetic datasets. 
 & $\mathcal{R}$ealistic, $\mathcal{S}$table \\
 \hline
 \ref{g7} & Sample data repeatedly under multiple data-generating processes, especially those that illustrate your methods' shortcomings.  
 & $\mathcal{S}$table, $\mathcal{T}$ransparent \\
 \hline\hline
 \multicolumn{3}{||l||}{\textbf{Preparation Steps (Methods)}} \\ 
 \hline
 \ref{g8} & Outline intended modeling procedure(s), including computational specifications as well as model forms and parameters. 
 & $\mathcal{T}$ransparent, $\mathcal{M}$odular \\
 \hline
 \ref{g9} & Include both commonly used methods and state-of-the-art competitors as comparative baselines. 
 & $\mathcal{I}$ntuitive, $\mathcal{E}$fficient \\
 \hline
 \ref{g10} & Select multiple performance metrics appropriate to the methods under comparison and their intended deployment. 
 & $\mathcal{S}$table, $\mathcal{R}$ealistic \\
 \hline\hline
 \multicolumn{3}{||l||}{\textbf{Staging (Reporting)}} \\ 
 \hline
 \ref{g11} & Keep displays -- whether visual or textual -- as simple, transparent, and digestible as possible for your intended audience. 
 & $\mathcal{I}$ntuitive, $\mathcal{T}$ransparent \\
 \hline
 \ref{g12} & Employ a diverse set of statistical summaries and visualizations to offer concrete evidence from multiple perspectives. 
 & $\mathcal{S}$table, $\mathcal{T}$ransparent \\
 \hline
 \ref{g13} & Do not mislead your audience, nor draw conclusions beyond the scope of your study's evidential support. & $\mathcal{T}$ransparent, $\mathcal{S}$table \\
 \hline
\end{tabular}
}
\end{center}
\caption{Thirteen guidelines for designing a simulation study with \texorpdfstring{$\mathcal{MERITS}$}{MERITS}.}\label{tab:guidelines}
\end{table}

\subsection{Recipe Header (Problem Statement)} \label{sec:header}

A recipe's header specifies its goal and outlook: respectively, (a) the identity and provenance of the dish -- for example, ``Grandma's apple pie'' -- and (b) expected yield and time to completion.  These same elements comprise a research problem statement, which describes (a) the study goal, with relevant context or motivating antecedents, and (b) the expected results, otherwise known as research hypotheses.  In each setting, these elements drive the project and dictate many subsequent decisions. 

\vspace{-0.3cm}
\begin{center}
\rule{300pt}{0.5pt}
\end{center}
\vspace{-0.5cm}

\begin{formal}
\begin{enumerate}[resume=guidelines,label=\roman*.,ref=\roman*]
    \item Describe as concretely as possible the research question(s) your study seeks to answer. [Principal $\mathcal{MERITS}$: $\mathcal{T}$ransparent, $\mathcal{I}$ntuitive] \label{g2}
\end{enumerate}
\end{formal}

\noindent Both chef and data scientist need a clear vision of their project goals to plan a successful approach.  In cooking, the intended dish dictates the choice of recipe and any impromptu decisions the chef makes in its pursuit.  Likewise, the problem statement drives a study and governs all decisions in its design and implementation.  Although we try to minimize \textit{ad hoc} decision-making (per guideline~\ref{g1}), we cannot foresee every contingency, and scientific progress requires a feedback loop across simulation, theory, and empirical investigation; research agendas evolve in response to new information.  Thus, in either setting, we must define our goals upfront [$\mathcal{T}$] in as precise and intelligible a manner as possible [$\mathcal{I}$], to ensure that the endeavor stays true to its purpose and sensible to its audience.

\begin{formal}
\begin{enumerate}[resume=guidelines,label=\roman*.,ref=\roman*]
    \item Connect your study to reality, while acknowledging its limitations honestly. [$\mathcal{R}$ealistic, $\mathcal{T}$ransparent] \label{g3}
\end{enumerate}
\end{formal}

\noindent One potential source of satisfaction in cooking and eating a meal grows from the sense of community and tradition that went into its creation, which promotes a feeling of cultural connection.  Likewise, an effective simulation gains relevance by connecting to a broader context.  Even mathematical or theoretical conjectures often are inspired by real-world problems or data.  Understanding a simulation study's motivation and impact requires some appreciation of the underlying domain problem; accordingly, the problem statement should draw connections between the simulation goals and the target of study [$\mathcal{R}$].

\medskip Often computational experiments must work with simplifications of the complex real-world phenomena they seek to describe, either for tractability or because (by the nature of research) we understand the target system imperfectly.  But a study's conclusions are only as actionable as they are realistic.  A study founded on implausible assumptions is at best irrelevant, and at worst, actively dangerous, if it promotes a misleading view of reality.  We should therefore assess and articulate upfront any limitations on interpreting our simulation's results with respect to the empirical world [$\mathcal{T}$].  Our resource constraints and imperfect knowledge might preclude solving every limitation, but reporting known imperfections aids our audience in grasping the scope and external validity of our conclusions.

\begin{formal}
\begin{enumerate}[resume=guidelines,label=\roman*.,ref=\roman*]
    \item Recognize and accommodate the wider research context for your study (\textit{e.g.}, domain knowledge or body of literature). [$\mathcal{E}$fficient, $\mathcal{R}$ealistic] \label{g4}
\end{enumerate}
\end{formal}

\noindent Seldom is there one `correct' way to prepare a dish or approach a research question.  Supposing we eliminate the flawed strategies, our choice among viable options often boils down to perspective or context.  Moreover, just as multiple dishes comprise a meal, so each simulation study contributes incrementally to communal knowledge:  We should design simulation studies with a holistic awareness of the existing literature, ensuring comparability with preceding studies while avoiding cumbersome redundancies [$\mathcal{E}$] -- akin to a chef designing new courses to complement an existing menu.  We require not only a thorough understanding of our own goals, but also of our antecedents: How have our predecessors undertaken similar tasks?  How did they succeed or fail, what did we learn from their experiences, and how does our endeavor complement or improve upon its precursors?  Of course, maintaining the fidelity of our study to the project goals and to reality is paramount: If obeying precedent would mean using an unsuitable (\textit{e.g.}, unrealistic, unreliable, or otherwise misleading) method, then we should defy convention in the interest of reliable conclusions [$\mathcal{R}$].

\vspace{-0.3cm}
\begin{center}
\rule{300pt}{0.5pt}
\end{center}
\vspace{-0.2cm}

\noindent \textbf{Putting it all together:} The problem statement of a simulation study is its foundation, and articulating it is perhaps the most critical phase of planning.  However, one needs not write a tome: The most-effective problem statements are punchy and concise.  Take the following excerpt from \cite{Hooker}, in which study the authors perform simple simulations to illustrate the limitations of existing variable-importance measures for interpreting black-box prediction algorithms:

\begin{quote}
This paper reviews methods based on permuting feature values or otherwise investigating the effect that changing the values of a feature has on predictions [...] \textit{When features in the training set exhibit statistical dependence, permute-and-predict methods can be highly misleading when applied to the original model.}
\end{quote}

\noindent The paper abstract and first paragraphs of the introduction elaborate on this statement, describing a give-and-take between the utility of black-box prediction algorithms and our imperfect understanding of their inner workings; the use of `permute-and-predict' methods to scry into such black boxes; and the authors' conviction that extant permutation methods do not produce reliable, actionable interpretations.  By the nature of their appearing in a published article, these statements are written more definitively than any problem statement the authors might have drafted upfront.  Nonetheless, the paper manifests many features of a strong problem statement: a clear statement of the research agenda (guideline \ref{g2}); an awareness of the relevant literature and antecedents (\ref{g4}); and the necessary burden of additional modeling required to support more-informative interpretive metrics (\ref{g3}).

\medskip A strong problem statement, as in the aforementioned paper, communicates the authors' purpose and strategy for achieving it, and provides a strong foundation for designing a study with $\mathcal{MERITS}$.  Subsequent design choices propagate from the problem statement.  Documenting it clearly and candidly, with explicit connections to the motivating problem or study system [$\mathcal{R}$], supports not only study reproducibility/replicability [$\mathcal{S}$], external validity [$\mathcal{S}$], and comprehension [$\mathcal{I}$] -- but especially transparency [$\mathcal{T}$].  If we are not comfortable preparing our dish in front of the diner, then we ought not feel comfortable in serving it.

\subsection{List of Ingredients (Data)} \label{sec:ingredients}

In cooking, the quality of our ingredients informs the taste and texture of our final dish, with subpar inputs generally affording subpar outputs.  Ingredient quality in turn depends on the manner of their growth and harvesting; handling in transit; and freshness upon arrival. This view extends naturally to the raw ingredients of simulation: the data or data-generating processes (DGPs).  The caliber of our conclusions depends on the fidelity of our data to the target of study (hence the adage, ``garbage in, garbage out'').  Such fidelity in turn depends on the manner of their selection and collection (growth and harvest); credibility of cleaning and partitioning processes (handling); and relevance to the target system (freshness).  We favor organic ingredients -- data close to nature -- over canned ingredients like the so-called ``toy'' examples often chosen for ease of simulation.

\vspace{-0.3cm}
\begin{center}
\rule{300pt}{0.5pt}
\end{center}
\vspace{-0.5cm}

\begin{formal}
\begin{enumerate}[resume=guidelines,label=\roman*.,ref=\roman*]
    \item Test methods in a (data) context as reminiscent as possible of their intended deployment.   [$\mathcal{I}$ntuitive, $\mathcal{S}$table] \label{g5}
\end{enumerate}
\end{formal}

\noindent The data we use to assess or refine a method should resemble those on which we will deploy it [$\mathcal{I}$].  Baking a cake looks very different from baking a chicken; we cannot perform the same procedure and expect a happy result in both cases, but rather, must tailor our technique to our ingredients.  Likewise, methods intended for applications with small samples or low signal-to-noise ratios (\textit{e.g.}, many clinical or genomics contexts) should not be studied solely on data with ample observations and strong signals; methods for continuous data might err when applied to integer counts. Confidence in a method can be only as strong as our confidence in the relevance of data used to develop it.  By extension, if we wish to deploy a technique in a new context (as often is the case with statistical tools), we must understand how the new setting defies or admits of our existing knowledge of the technique [$\mathcal{S}$].  A method might accommodate multiple types of data, but its appropriateness should be considered before being used ``off label.''  And if we hope the method will suit many settings, then our study should encompass many data types to demonstrate this capability.

\begin{formal}
\begin{enumerate}[resume=guidelines,label=\roman*.,ref=\roman*]
    \item Incorporate real-world data and/or accepted scientific theory in generating synthetic datasets.  [$\mathcal{R}$ealistic, $\mathcal{S}$table] \label{g6}
\end{enumerate}
\end{formal}

\noindent Refining a technique requires repetition and practice.  In cooking, we might hone our skills by practicing on subpar ingredients, reserving our best specimens for preparing a dish in earnest.  But if our practice ingredients are too dissimilar to the proper ones -- for example, if they have a very different texture from being under- or over-ripe -- then we might not be able to use the same techniques on both.  In Data Science, we can think of real-world data as our ``best specimens'' and of synthetic data as our practice ingredients.  We assess models using synthetic data with a known truth baked in for the methods to discern; we then deploy our preferred model(s) on real-world data.  Confidence in the realism [$\mathcal{R}$] and generalizability [$\mathcal{S}$] of a model requires that the synthetic data resemble real-world data.

\medskip Canned ingredients offer convenience and reliability, and toy examples are not without virtue.  The conventional repertoire of ``canned'' DGPs (\textit{e.g.}, Gaussian data with independent errors) offers a benchmark for comparing across studies and thus for assessing the relative performance of methods.  But for external validity, we recommend supplementing conventional DGPs with `natural' DGPs that resemble reality.  Specifically, we recommend incorporating real-world data into the generation of simulated datasets and using predictability checks to assess the quality of simulated features.  For example, we might generate synthetic phenotypes (\textit{e.g.}, disease incidence) using linear combinations of gene-expression data, injected with plausible random errors.  We incorporate real-world complexities (gene correlation structure) that might affect model performance -- approaching our practice ingredients to the real ones -- to prevent unforeseen complications or misleading results when deploying the method in earnest.  Incorporating such complexities through sampling allows us to learn from the underlying data structure without quantifying it.

\begin{formal}
\begin{enumerate}[resume=guidelines,label=\roman*.,ref=\roman*]
    \item Sample data repeatedly (using distinct random seeds) under multiple DGPs, especially those that illustrate your methods' shortcomings.  [$\mathcal{S}$table, $\mathcal{T}$ransparent]\label{g7}
\end{enumerate}
\end{formal}

\noindent Different cooking techniques suit different ingredients, and the best dishes arise from knowing the strengths and weaknesses of each pairing.  Similarly, gaining a robust understanding of a statistical method requires exploring its performance in a variety of realistic contexts.  To that end, DGPs should be varied (\textit{e.g.}, across sample sizes, parameter values, distributions of randomness)\footnote{Note the governing role of the problem statement: These and other variations may be more or less pertinent to one's research agenda and their inclusion or omission should be considered and justified.} and should include scenarios that might prove unflattering to the methods under consideration.  We also recommend repeating the data-generation process and the experiment as a whole under various random seeds to mitigate the role of luck.  These efforts combine to illustrate the stability of results across reasonable perturbations to one's analysis pipeline -- not only encouraging a well-rounded view of each individual method [$\mathcal{S}$], but also allowing researchers to compare methods (within and across studies) without fear that the data were cherry-picked to support a particular conclusion.

\medskip We recognize some tension between the goal of exploring the problem space adequately and that of creating an efficient study design.  We encourage prioritizing DGPs that might clarify a method's limitations [$\mathcal{T}$], such as those with low signal or that violate the model assumptions.  Exploring scenarios that might tax or break a method allows us to recognize and accommodate its restrictions.  We can adapt more readily to quirks of our data, loosely akin to a chef's ability to substitute for an out-of-season ingredient while preserving the integrity of a dish.  And while the individual researcher undoubtedly benefits from a better understanding of a method's capabilities, our communal knowledge benefits even more because we lose less time relearning redundant lessons and instead can focus our energies on developing or refining methods to fill the gaps.

\vspace{-0.3cm}
\begin{center}
\rule{300pt}{0.5pt}
\end{center}
\vspace{-0.2cm}

\noindent \textbf{Putting it all together:} In Data Science as in cooking, the quality of our ingredients informs the quality of our results.  \cite{ZIFA} offers a prime example of a study whose authors attended to the suitability of their ingredients.  This paper describes the stress-testing of a method, Zero-Inflated Factor Analysis (ZIFA), for dimensionality reduction of single-cell gene expression data with a high prevalence of dropouts (hence, an inflated number of zero counts).  

\medskip The study comprises two parts.  In the first, the authors generated a battery of datasets designed to test ZIFA's robustness to model misspecification, assessing model performance when acting on data simulated under various ``noise levels, dropout rates, number of latent dimensions and number of genes.''  The datasets shared a generative model and base simulation settings, with each simulation episode varying one of the base parameters to learn the effect of that particular perturbation on method performance.  In the authors' words, this approach ``was not intended to truly reflect actual real world data characteristics but to establish, when all other modeling assumptions are true, the impact of dropout events.''  The second part of the study employed what we would call ``real-data inspired'' simulation: For each of four empirical gene-expression datasets, the authors sampled genes on which to fit their method and (lacking, in this context, a known truth for the latent space) compared the resulting posterior predictive count distributions to the observed data distribution. 

\medskip In sum, \cite{ZIFA} deliberately tested ZIFA on appropriate and varied data (guideline \ref{g5}) generated using processes inspired by and involving real-world data (\ref{g6}) under assorted conditions, including those with the potential to illuminate the method's limitations (\ref{g7}).  The resulting study design touches on all six $\mathcal{MERITS}$: Their choices of DGP suited the study goals and audience [$\mathcal{I}$], and the design efficiently [$\mathcal{E}$] (a) incorporated real-world complexities to ensure the study's relevance to reality [$\mathcal{R}$] and (b) explored various scenarios [$\mathcal{M}$], especially those challenging to the method [$\mathcal{T}$], to ensure a robust understanding of its capabilities [$\mathcal{S}$].  Only with a thorough understanding of the interplay between techniques and ingredients can we create the finest products, and \cite{ZIFA} gave readers a thorough understanding of the technique they developed (ZIFA) to operate on their chosen flavor of ingredient (zero-inflated gene-expression data). 

\subsection{Preparation Steps (Methods)} \label{sec:prep}

Culinary school teaches a repertoire of cooking techniques for use whenever a particular dish demands.  Analogously, the data scientist's repertoire includes various methods for use as needed to answer a particular research question.  We choose our methods to suit both the intended dish (research question) and available ingredients (data).

\vspace{-0.3cm}
\begin{center}
\rule{300pt}{0.5pt}
\end{center}
\vspace{-0.5cm}

\begin{formal}
\begin{enumerate}[resume=guidelines,label=\roman*.,ref=\roman*]
    \item Outline intended modeling procedure(s), including computational specifications as well as model forms and parameters. [$\mathcal{T}$ransparent, $\mathcal{M}$odular] \label{g8}
\end{enumerate}
\end{formal}

\noindent As the intended dish dictates a chef's choice of equipment and techniques, so should the problem statement govern a data scientist's choice of methodology.  These details include statistical paraphernalia -- model forms and parameters -- but also the overall analytical pipeline and implementation: code architecture and dependencies; strategies for random-number generation and parallelization; and records to keep by simulation replicate (\textit{e.g.}, success declaration in a simulated trial) and overall (\textit{e.g.}, trial power).  Such decisions must suit our ingredients and \textit{vice versa}.  Code should be documented and logically partitioned such that we can add or exchange methods (or DGPs) without revamping the entire pipeline [$\mathcal{M}$].  Such modular construction allows a study to evolve (\textit{e.g.}, for new data, or comparison against untried methods), thereby improving its longevity and usefulness to the field. 

\medskip Planning these particulars upfront minimizes inadvertent biases and (recalling the \textit{mise en place}) mitigates chaos in implementation.  Doing so also encourages us to think about the overall scope of the design, to address crucial elements while keeping study execution manageable.  Insofar as we cannot anticipate all contingencies, we record any \textit{post hoc} decisions made during study execution for transparency and to inform future work [$\mathcal{T}$].

\begin{formal}
\begin{enumerate}[resume=guidelines,label=\roman*.,ref=\roman*]
    \item Include both commonly used methods and state-of-the-art competitors as comparative baselines. [$\mathcal{I}$ntuitive, $\mathcal{E}$fficient] \label{g9}
\end{enumerate}
\end{formal}

\noindent Propagating a new technique means persuading potential users that its benefits justify the time and effort required for uptake. To that end, simulation studies should include both conventional and cutting-edge techniques as comparative baselines so that readers can interpret the results relative to familiar methods as well as industry standards [$\mathcal{I}$].  Such intellectual footholds help the audience digest the study, rendering it less alienating and boosting confidence in its quality and fairness.  This strategy also might mean favoring methods with existing implementations, given that practitioners often gravitate to methods with existing software/repositories.  But precedent alone does not justify including a method.  No chef would choose a blunt knife over a sharp one -- and if another method has been shown superior, then comparing against an inferior procedure adds unnecessary bulk to both our code and our results table [$\mathcal{E}$].  We therefore recommend excluding methods once superseded, to avoid both the paper-tiger fallacy of comparing against a trivially easy opponent and the generation of cumbersome results tables.

\begin{formal}
\begin{enumerate}[resume=guidelines,label=\roman*.,ref=\roman*]
    \item Select multiple performance metrics appropriate to the methods under comparison and their intended deployment. [$\mathcal{S}$table, $\mathcal{R}$ealistic] \label{g10}
\end{enumerate}
\end{formal}

\noindent Evaluating a prepared dish is a holistic process, incorporating simultaneous judgements of texture, taste, and aesthetics.  It moreover depends on context, including the dish's function (appetizer, main course, or dessert) and the setting (fast-food joint or fancy restaurant).  Similarly, in Data Science, we recommend evaluating models using multiple metrics (\textit{e.g.}, mean squared error \textit{versus} mean absolute error) across multiple dimensions (\textit{e.g.}, computational efficiency as well as fit).  Diversifying our evaluation metrics affords a multifaceted understanding of a method's strengths and weaknesses, much as varying our DGPs highlights its suitability to different applications [$\mathcal{S}$].  As always, our choices should reflect the study goals and the reality of how our methods will be deployed in practice [$\mathcal{R}$].  For example, we can contrast the prediction accuracy of a linear model and ``black-box'' predictive algorithm, but comparing the internal structure of the two models is difficult if not impossible.  If practitioners care about prediction but not interpretability or model complexity, then our inability to compare models on the latter dimensions is admissible.

\vspace{-0.3cm}
\begin{center}
\rule{300pt}{0.5pt}
\end{center}
\vspace{-0.2cm}

\noindent \textbf{Putting it all together:} A similar logic governs choosing a study's `equipment' and its `ingredients'.  We wish to mitigate bias and produce actionable, reliable results [$\mathcal{R}$, $\mathcal{S}$].  Once again, this goal requires a trade-off between plumbing the problem space and designing a study both executable and digestible [$\mathcal{E}$].  Efficient simulations minimize bloated code (for ease of preparation) [$\mathcal{M}$] and results (for ease of digestion) [$\mathcal{I}$].  Our problem statement informs our choices of models to fit and performance metrics to evaluate, among other specifications, all of which we report upfront in the project documentation [$\mathcal{T}$].

\medskip Consider a stimulation study by \cite{Wang23}, which entailed a comparative analysis of 21 dimension-reduction (DR) methods acting on single-cell flow cytometry data.  The authors describe their methods and rationale in admirable detail (guideline \ref{g8}).  They select deliberately for a combination of ``general-purpose'' DR methods and those ``developed specifically for single cell'' data; moreover, they include as an intellectual benchmark the two DR methods preeminent in everyday use among practitioners (tSNE and UMAP) (\ref{g9}).  

\medskip The authors attend with similar care to their choice of evaluation metrics.  In their own words, ``for scoring the accuracy of the DR methods, we chose a total of 16 metrics in four main categories,\footnote{Namely, ``(1) global structure preservation, (2) local structure preservation, (3) downstream analysis performance, and (4) concordance of DR results with matched scRNA data.''} characterizing different aspects of the performance'' (\ref{g10}).  Beyond assessing the methods' statistical capacities, they report on pragmatic features of each method, including ``scalability with respect to the number of cells and protein markers, stability of the DR after re-sampling the datasets and also parameter tuning; and the usability of the tools in terms of software and documentation.''  These various metrics unite to provide a multi-faceted understanding of available methods' relative strengths and weaknesses -- a necessity in this setting, where (the authors conclude) different DR methods do not dominate but rather complement one another, and ``the choice of method should depend on the underlying data structure and the analytical needs.''

\subsection{Staging (Reporting)} \label{sec:staging}

Staging a dish is the most subjective phase of its preparation, in that different diners (or food critics!) have different aesthetic tastes, and thus may diverge in whether and how they welcome a given culinary presentation.  The ``best'' method of communicating simulation results depends similarly on the background and perspective of the intended audience. 

\vspace{-0.3cm}
\begin{center}
\rule{300pt}{0.5pt}
\end{center}
\vspace{-0.5cm}

\begin{formal}
\begin{enumerate}[resume=guidelines,label=\roman*.,ref=\roman*]
    \item Keep displays -- whether visual or textual -- as simple, transparent, and digestible as possible for your intended audience. [$\mathcal{I}$ntuitive, $\mathcal{T}$ransparent] \label{g11}
\end{enumerate}
\end{formal}

\noindent The staging of a finished entr\'ee should enhance rather than detract from its enjoyment, adding to the dish's visual appeal without damaging its taste, texture, or ease of consumption.  In Data Science, the same credo argues for presenting results in a digestible format, from which readers can draw accurate conclusions comfortably [$\mathcal{I}$].  Ideally we present results in a manner intelligible to a broad audience, and some principles are suitably ubiquitous: employing italic or bold text \textit{sparingly} for emphasis and to guide the reader's eye over tables and figures; choosing accessible color palettes for visualizations; balancing mathematical notation against written explanations; and interpreting results candidly and concisely [$\mathcal{T}$].  However, different readers may consider different presentations to be the pinnacle of ``simple, transparent, and digestible'' -- much as a patron at a family restaurant might prefer a staging more approachable and less artistic than a food critic at an upscale establishment.  We should prioritize our target audience in deciding, for example, the balance between visual and tabular presentations; how much supplementary information or interpretation to include; and the appropriate level of literary formality. 

\begin{formal}
\begin{enumerate}[resume=guidelines,label=\roman*.,ref=\roman*]
    \item Employ a diverse set of statistical summaries and visualizations to offer concrete evidence from multiple perspectives. [$\mathcal{S}$table, $\mathcal{T}$ransparent] \label{g12}
\end{enumerate}
\end{formal}

\noindent An epicure experiences an entr\'ee using multiple senses (\textit{e.g.}, sight, smell, taste).  Engaging with the subtleties of a complex simulation study requires a similar versatility of approach.  Likely no single table or figure captures all pertinent information elegantly, especially if we have explored a variety of DGPs, methods, and performance metrics as per the preceding guidelines.  A method's character (good, bad, or mixed) may become clear only across multiple vantage points suited to different modes of communication.  And as just discussed, different audience members may learn best from different presentations of results.  Thus, providing multiple perspectives affords the most-robust understanding to the most-comprehensive audience, while also preventing the cherry-picking of results that flatter any one particular method [$\mathcal{T}$].  We recommend providing a judicious handful of summaries and graphics: varied enough to appeal to a broad audience and to provide a nuanced picture of the study results [$\mathcal{S}$], but few enough to avoid overwhelming readers.

\begin{formal}
\begin{enumerate}[resume=guidelines,label=\roman*.,ref=\roman*]
    \item Do not mislead your audience, nor draw conclusions beyond the scope of your study's evidential support. [$\mathcal{T}$ransparent, $\mathcal{S}$table] \label{g13}
\end{enumerate}
\end{formal}

\noindent Just as peers in a culinary school might compare and critique one another's dishes, so do data scientists learn from one another's work -- the mishaps as well as the successes.  To that end, we should represent each method's capabilities fully and accurately, not excepting the scenarios under which it falls short, in the interest of both intellectual honesty and community growth [$\mathcal{T}$].  Figures should adopt reasonable and appropriate axis scales and include error bars when possible; written summaries should report results frankly, without hyperbole or dissembling.  Authors' discussions should rate their confidence in and external validity of conclusions based on such factors as the perturbations explored, number of replicates executed, and stability of results across variations or random seeds [$\mathcal{S}$]. Overstating or skewing one's results undermines readers' confidence in the work, suggesting that the responsible parties either did not perform a sufficiently rigorous study or else distorted the outcome intentionally -- and without the audience's confidence, one's legacy (whether in the statistical or the culinary sciences) must be destined for the rubbish bin.

\vspace{-0.3cm}
\begin{center}
\rule{300pt}{0.5pt}
\end{center}
\vspace{-0.2cm}

\noindent \textbf{Putting it all together:} We offer fairly broad recommendations for staging the results of a simulation study because such decisions depend on the nature of the investigation and of the target audience.\footnote{Additional resources can be found in our literature review (Section~\ref{sec:disc} and Appendix~\ref{sec:further}).}  \cite{Huang25} provide a nice selection of summaries for comparing methods across an array of procedures, simulation variants, and performance metrics.\footnote{For transparency, we note that the second author is also a co-author on the manuscript you are reading.}  This study explored the utility of a new method for identifying meaningful subgroups of a population (\textit{i.e.}, patients who benefit notably well or poorly from a given intervention) as `distilled' from black-box algorithms for learning heterogeneous treatment effects.  The authors provide a variety of visual and tabular summaries (particularly in the Supporting Information), including, but not limited to, line plots of performance metrics; heatmaps for the proportion of simulations in which, under various conditions, each covariate was identified as influential; stacked bar charts to illustrate stability in feature selection across simulation settings; and tables of effect estimates.  Perhaps most importantly, the authors prioritized reproducibility and transparency by documenting and publishing all code and results in an open-source repository. Such clarity [$\mathcal{I}$] and transparency [$\mathcal{T}$] helps ensure that reliable and actionable new knowledge [$\mathcal{R}, \mathcal{S}$] reaches a wide audience in a format easily digested and disseminated. 

\subsection{Coda: Tasting as we go (execution and iteration)} \label{sec:iteration}

Herein we argue for thorough planning prior to implementation.  Nonetheless, we recognize that executing a high-quality simulation study, like preparing a superlative entr\'ee, requires learning from experience.  A study with fully predictable results is superfluous, and while reusing real-world data can give rise to significant biases, refining a simulation is less treacherous (provided we maintain transparency and protect against misleading results).  Thus, we acknowledge an implied sequel to planning the simulation recipe, in which the data scientist (a) executes the study and (b) responds to its results, perhaps by refining the original recipe or designing further experiments.  Because the nature of a study's evolution depends greatly on its context and history, herein we do not dwell on these stages except to note that the $\mathcal{MERITS}$ of a high-quality simulation should guide study \emph{revision} as well as \emph{design}.  Moreover, a study's evolution should align with the problem statement (like tweaking a recipe to improve the dish) rather than changing targets to accommodate the study's shortcomings.  Most importantly, the study record should justify any \textit{ad hoc} decisions, for both reproducibility and context in interpreting results.

\section{A page from our cookbook: Case Study}\label{sec:casestudy}

Having introduced our PCS primer, we now consolidate and illustrate these ideas by dissecting a methodological simulation study.\footnote{Appendix~\ref{sec:VRL} analogously autopsies a non-methodological (emululatory) study in the social sciences.}  \cite{iRF} introduced the iterative random forest (iRF) algorithm for discovering interactions in high-dimensional data.\footnote{Specifically, epistatic interactions driving gene expression in high-dimensional, labelled \textit{omics} data.}  Development drew extensively on simulations to identify settings in which the method performed well or poorly, and to compare against other methods.  Performance criteria included computational efficiency as well as the validity of interactions learned from both synthetic and empirical (ChIP-seq) data.  The authors planned the study upfront (per guideline~\ref{g1}), from goals and scope to consequent choices of data (generation), methods, and research products.

\medskip In this case study, the authors leverage simulation to understand their method's strengths and limitations in data contexts reminiscent of its intended use by practitioners.  However, we pause to note that simulation acts both as an exploratory tool and as a \textit{complement to} to rather than a \textit{replacement for} derivation of the method's theoretical properties.  For an example of a paper blending theoretical and simulation-based method scrutiny -- a topic beyond the scope of this manuscript -- we refer the interested reader to \cite{LSSFind}.  In this sequel to \cite{iRF}, the authors employ both simulation and analytical derivation to evaluate the theoretical properties of tree-based interaction discovery.\footnote{We discuss the simulation part of the sequel study in Appendix~\ref{sec:LSSFind}.}

\subsection*{Recipe Components}

\begin{enumerate}
\item[A.] \textbf{Recipe Header (Problem Statement)}: \textit{What are we trying to produce?} \\ \cite{iRF} sought to evaluate iRF in settings representative of its intended \textit{omics} applications (\ref{g2}), comparing against existing methods in terms of computational efficiency and ability to detect interactions. The authors describe the general problem of interaction discovery and its empirical relevance to learning genetic activity from ChIP-seq data; they acknowledge that validating computational `discoveries' is difficult due to a dearth of empirical evidence for genetic interactions (\ref{g3}).  The paper describes existing methods and compares iRF against two such (\ref{g4}).     

\item[B.] \textbf{List of Ingredients (Data)}: \textit{What should go into it?} \\ The iRF simulation study employed feature matrices built from either real-world ChIP-seq data or synthetic data designed to emulate gene dynamics (\ref{g6}).  In either case, the authors seek to maintain the fidelity of simulated data to reality.  Their simulations on empirical data most-directly resemble the intended usage of iRF for biological discovery (\ref{g5}).  For synthetic datasets, the authors generated feature matrices with various properties potentially difficult for iRF (\textit{e.g.}, feature correlation, number of features, mixture distributions across features) to assess their individual effects on method performance (\ref{g7}). Response variables then were derived from these feature matrices \textit{via} a generative model whose functional form mimics our understanding of the biological interactions regulating genomic activity, in that:

\begin{itemize}
    \item Individual features are thresholded prior to consolidation.  This transformation reflects the understanding that cells behave differently when inputs exceed specific levels.  Such thresholding dynamics are thought to influence the cells' differential development and functioning \citep{wolpert1969positional}.
    \item Thresholded features are consolidated using linear combinations of Boolean functions.  This mechanism reflects the dynamics of cell complexes and cooperative/competitive binding (as reviewed in \citealt{spitz2012transcription}).
\end{itemize}

\item[] The so-called ``local spiky sparse'' (LSS) model uniting these two considerations embodies our goal of suiting one's DGPs to the problem statement and related domain knowledge.  Other types of data (\textit{i.e.}, non-binary) would require similarly thoughtful design of generative models that both resemble relevant empirical data and afford suitably generalizable conclusions.  In settings with a broad research agenda, we recommend simulating across a variety of data scenarios to assess a method's robustness to -- and, by extension, the stability of results across -- data perturbations.

\item[C.] \textbf{Preparation Steps (Methods)}: \textit{What steps are required?} \\ The primary method under consideration for this project was, of course, iRF itself.  This method builds on both random forests (RFs) and random intersection trees (RITs): It incorporates iteratively re-weighted RFs to map continuous data to sets of key features, and RITs to search for influential combinations thereof. In addition to the standard tuning parameters for those two methods, iRF adds a parameter $K$ specifying the number of iterations.  Simulations compared iRF against the conventional RF and state-of-the-art alternative Rulefit 3 (\ref{g9}).  The full iRF algorithm, antecedents, competitors, and implementation all are described in~\cite{iRF} (\ref{g8}).  Algorithm performance was evaluated in terms of computational efficiency and across three different perspectives on interaction discovery: ranking quality, true-positive rate, and false-positive rate (\ref{g10}). 

\item[D.] \textbf{Staging (Reporting)}: \textit{How will the finished product look if done right?} \\ The authors reported the results of their iRF study using a combination of flow-charts describing the algorithmic procedure; narrative text; line graphs of precision-recall curves under different values of $K$; scatterplots of iRF-generated interaction-quality scores; and surface plots and heatmaps of coincident biomarker expression (\ref{g12}). For each simulation episode, they provide a consistent set of summary figures to support comparisons across data settings (\ref{g11}). For reproducibility and transparency, the authors provide all datasets in the paper’s Supporting Information and published all code, documentation, and data on the research-sharing service, Zenodo (\ref{g13}).
\end{enumerate}

\subsection*{\texorpdfstring{$\mathcal{MERITS}$}{MERITS} and Shortfalls}

\begin{itemize}
\item $\mathcal{M}$\textit{odular}: The authors of the iRF study composed their response models from simple Boolean ({\tt AND}, {\tt OR}, {\tt XOR}) rules that could be applied to any data matrix, thereby allowing the same model to operate on both simulated and real-world datasets.  They further wrote the modeling code flexibly to accommodate different strengths of feature interaction, types of response, and choices of response threshold.  Thus, they were able to recycle the code architecture to evaluate their methods under these various data settings. The response models did not allow for facile composition of individual {\tt AND}, {\tt OR}, and {\tt XOR} rules into more-complex modeling rules, but the authors implemented this extension in a follow-up paper~\citep{siRF}.

\item $\mathcal{E}$\textit{fficient}: The authors considered a range of data settings to emulate real-world complications, including correlated features, variables outnumbering observations, low signal-to-noise ratios, and observation-dependent noise.  The final study comprised one simulation examining each of these complications (a divide-and-conquer format) for clarity and computational efficiency.  The efficiency of iRF itself was limited by its reliance on the {\tt R} package {\tt randomForest}, but the authors later updated {\tt iRF} to work with an optimized random-forest algorithm available from the package {\tt ranger}.

\item $\mathcal{R}$\textit{ealistic}: The authors developed iRF to discover biomolecular interactions in \textit{omics} data.  To understand the performance of iRF in this setting, they (a) generated response values from Boolean-type rules intended to reflect biological dichotomization; (b) evaluated the performance of iRF using real-world ChIP-seq data; and (c) evaluated iRF under a range of common data complications (see previous bullet). 

\item $\mathcal{I}$\textit{ntuitive}: For clarity of scope, the authors adopted a divide-and-conquer format comprising several simulations, each devoted to a specific data setting.  In choosing parameters, they sought to evaluate a wide enough range of settings to afford a robust view of method performance while keeping results concise for clear interpretation.  The authors considered three performance metrics to quantify different facets of ``successful interaction recovery'': recovery (\textit{i.e.}, recall) rate; false-positive rate; and area under the ROC curve, with recovered interactions ranked by a quality score.  Although tastes might vary, the authors deliberately chose to present a holistic picture of performance rather than trying to define a singular performance metric. 

\item $\mathcal{T}$\textit{ransparent}: The authors published all simulation code and data on the research-sharing service, Zenodo.  They documented the code thoroughly to aid future users in understanding its purpose and operations.  They did not provide the simulations as integrated/dynamic documents (in Jupyter or {\tt R} markdown), but described the study procedure in the publication and provided the code separately.

\item $\mathcal{S}$\textit{table}: To support future users of iRF, the authors provide an implementation of the algorithm as an {\tt R} package, {\tt iRF}.  They employed this implementation under a wide range of response models, parameters, and feature matrices to evaluate the stability of iRF performance across many settings.  For reproducibility, they set and recorded random seeds for consistent random-number generation, and performed the simulations on multiple operating systems to ensure stability across computing platforms.
\end{itemize}

\section{Simulation culture: Connections to the literature} \label{sec:disc}

Various researchers have advised the Data Science community on aspects of good simulation practice, but we found no unified guidance that seemed to have widespread acceptance among practitioners.  We now connect our PCS primer to various antecedents, partitioning our discussion (with admittedly fuzzy borders) across contributions to study conceptualization, design, and appraisal.  For readers with specific interests (\textit{e.g.}, empirical support, coding principles and software), we refer the reader to additional discussion in Appendix~\ref{sec:further}.

\subsection*{Conceptualization}

Our PCS primer takes its name from the Predictability-Computability-Stability (PCS) framework for veridical Data Science~\citep{Yu, Yu-Kumbier}. The PCS framework straddles the divide between study conceptualization and appraisal, as the authors recommend grounding statistical analyses with connections to the eponymous three pillars of Data Science.  \textit{Predictability} invokes prediction as a model ``reality check,'' and falls under our pursuit of $\mathcal{R}$ealism in translating between empirical research question and computational construct.  Both views of $\mathcal{S}$tability embed not only statistical variation, but also reasonable variation in human judgements, to promote a standard of reproducibility/replicability and external validity as in the empirical sciences. And our $\mathcal{M}$odularity and $\mathcal{E}$fficiency support the \textit{computability} of a simulation, which comprehends such concerns as data collection and storage, access and cleaning; and the design of algorithms that are tractable, efficient, and scalable.  Auxiliary to the PCS acronym, its authors recommend a high standard of ``PCS documentation''; we elevate this fourth, critical dimension to a more-equal footing through our principles of $\mathcal{I}$ntuitiveness and $\mathcal{T}$ransparency, which encourage clear and honest communication of a study and its results.  Our PCS primer moreover offers a conceptual scaffold and tangible guidelines to aid Data Science practitioners in designing simulations.

\medskip \cite{Donoho} argues that Data Science has accelerated rapidly with three pillars of ``frictionless reproducibility'' -- communal sharing of data (FR1), code (FR2), and competitive challenges (FR3) -- that support fast, grounded research proliferation. Donoho's narrative, while not limited to simulation, intertwines with our PCS-inspired $\mathcal{MERITS}$. All three pillars support research Computability, giving future researchers a stronger computational foundation on which to build. Shared data (FR1) underlie $\mathcal{R}$ealistic simulations, while $\mathcal{R}$ealistic synthetic data serve as benchmarks in competitive challenges (FR3)~\citep{Carvalho, Dorie}. Researchers can more easily adapt shared code (FR2) that is $\mathcal{M}$odular and $\mathcal{E}$fficient. And for these pillars to truly reduce friction in advancing human knowledge, all three must possess $\mathcal{I}$ntuitive, $\mathcal{T}$ransparent documentation for utility and user confidence. Overall, research products embodying the $\mathcal{MERITS}$ of good Data Science form more-reliable foundations for future work via FR1-3, and are more likely to promote $\mathcal{S}$table truths rather than conclusions that later prove false.

\medskip In a similar vein, \cite{Morris} recognized the poor reporting of published simulations and argued for better communication of study design and execution.  They suggest the mnemonic ADEMP (aims, DGPs, estimands, methods, and performance measures) to help practitioners articulate their designs; \cite{Boulesteix} revised ADEMP to replace `estimands' by `number of repetitions'. Our PCS primer presents these considerations in a more-narrative format \textit{via} three of our four simulation ``recipe'' components, with associated design guidelines: the problem statement (aims); data (and DGPs); and methods (including estimands, performance measures, and computational specifications).  We further recognize explicitly in our fourth component the importance of proper reporting, while \cite{Williams24} and \cite{Siepe} extend ADEMP to recommend foci for reporting on simulations in, respectively, ecology/evolution and psychology. 

\subsection*{Design}

The Systems Engineering and Computer Science Program (PESC) developed 20 guidelines for reporting simulations in software engineering, which we can interpret retroactively as design considerations~\citep{PESC}.  They, too, recommend documenting the domain problem, background knowledge and study context, and research goals (problem statement guidelines); truth scenarios (data); modelling details, validation efforts, and computational parameters (methods); and main findings and limitations (reporting).

\medskip A U.S. interagency working group sought to improve the credibility of computational modeling in healthcare by establishing ten rules for simulation.  These precepts encourage context-appropriate choices, ample documentation, and reproducibility -- but do so at varying levels of specificity, from coarse (``evaluate within contex'') to very fine (``use version control''). Although we agree with their rules, we believe that the PCS primer's steadier level of granularity and recipe-like scaffold will facilitate internalizing and implementing it.

\medskip \cite{Collin} offer recommendations in personalized medicine that stand out for encompassing the entirety of the Data Science ``life cycle'': here, \textit{in silico} modelling, empirical studies, and clinical practice.  In consequence, they offer few precepts for study design \textit{per se}, but share our emphasis on careful prespecification and thorough documentation, with assumptions and biases recorded explicitly and models validated using empirical data. 

\subsection*{Appraisal}

Various collections of `Ten Simple Rules' advise on a selection of related topics, including big data~\citep{Zook}, molecular simulations~\citep{Elofsson}, behavioral modelling~\citep{Wilson}, open-access science~\citep{Rule}, and principled simulation~\citep{Fogarty}.  We can view these contributions alternately as design guidelines or quality-criteria checklists.  From various perspectives, they promote deliberate, streamlined code design ($\mathcal{M}$odularity and $\mathcal{E}$fficiency); empirical validation of computational models ($\mathcal{R}$ealism); clear communication through story-telling ($\mathcal{I}$ntuitiveness); `auditability' and recognition of data limitations ($\mathcal{T}$ransparency); and reproducibility ($\mathcal{S}$tability).

\medskip For readers interested in responsible data use and algorithm accountability, \cite{FAT-ML} offer five desiderata: Responsibility, Explainability, Accuracy, Auditability, and Fairness.  These principles govern at a more-abstract level than ours, but are compatible in that simulations designed with the PCS primer to attain its $\mathcal{MERITS}$ indirectly (but more tangibly) promote these higher-level ideals.  For further reading, we recommend \cite{Ashurst} and \cite{REAL-ML}.  For granular discussions of computational reproducibility and coding recommendations -- of which there are many beyond the scope of our discussion -- we suggest \cite{Krafczyk} and \cite{factor12}. 

\section{Concluding remarks}

Simulations are ubiquitous in and invaluable to contemporary research in many disciplines, offering an efficient and (potentially) realistic training ground for models in various data contexts.  Yet this prevalence has fostered no standard criteria for a ``high-quality'' study, nor a curated guide to designing one.  Absent such guidance, novice data scientists evolve their own practices \textit{ad hoc}, informed by word of mouth or individual trial and error, both of which processes are inefficient and fallible. Herein we attempt to unite the field and encourage trustworthy, veridical Data Science \textit{via} our PCS primer for designing high-quality simulation ``recipes.''  Our $\mathcal{MERITS}$ of a strong simulation ($\mathcal{M}$odular, $\mathcal{E}$fficient, $\mathcal{R}$ealistic, $\mathcal{I}$ntuitive, $\mathcal{T}$ransparent, $\mathcal{S}$table), and guidelines in pursuit of them, seek to formalize lessons learned from community simulation in the natural and computational sciences.  We hope that this consolidated presentation of ideas will aid data scientists in teaching and adopting good practices easily and consistently, and mitigate dubious research practices.  

\medskip To a similar end, we have designed the {\tt R} package {\tt simChef} to facilitate implementing and reporting simulation studies with $\mathcal{MERITS}$.  Unlike existing simulation packages, {\tt simChef} supports Tidyverse-esque syntax for writing simulations and automatic generation of R Markdown documentation containing simulation results with the code to reproduce them.  Over repeated use, our primer would become second nature -- much as an experienced chef knows intuitively which flavors complement one another -- but good practices proliferate more effectively when adopting them is easier than the alternative.  Thus, with {\tt simChef}, we seek to lower the activation barrier to running and reporting high-quality simulations, allowing researchers to focus on substantive questions with fewer technical distractions.  We refer interested readers to the \href{https://github.com/Yu-Group/simChef}{Github repository} and manuscript~\citep{simChef}. 

\medskip We recognize that it may not be possible to satisfy all nuances of this framework in every context.  But, should the Data Science community choose to embrace our PCS primer, we hope that it will support a higher caliber of data-driven science; help mitigate the so-called reproducibility crisis; and even facilitate more-prosaic tasks like preparing research for publication.

\section{Disclosure statement}

The authors have no competing interests to declare.

\pagebreak

\bibliographystyle{agsm}
\bibliography{references}

\pagebreak

\appendix

\section{Appendix: LSSFind (theory-driven case study)}\label{sec:LSSFind}

\cite{LSSFind} sought to extend and validate the iRF project (Section~\ref{sec:casestudy}) by proving that an ``iRF-like'' method discovers `true' feature interactions.  The authors used a local spiky sparse (LSS) model to simulate interaction terms and tried to recover them using a simplified version of iRF, dubbed LSSFind.  The paper describes a largely theoretical evaluation of tree-based interaction discovery; herein we consider only the complementary simulation study.  The authors generated various datasets, including several that violated their modeling assumptions, to verify and stress-test their conclusions.  Because they worked primarily with a simplified algorithm, the authors also included a direct comparison of the simplified (LSSFind) and full (iRF) methods.  As per our primary guideline (\ref{g1}), the authors planned the study largely prior to implementation.  However, the auxiliary data-inspired simulations (discussed below) required a few iterations of learning from intermediate results.

\subsection*{Recipe Components}

\begin{enumerate}
\item[A.] \textbf{Recipe Header (Problem Statement)}: \textit{What are we trying to produce?} \\ \cite{LSSFind} sought to (a) recover from simulation their theoretical conclusions; (b) assess the breakdown of these conclusions under model-assumption violations; and (c) compare the simplified LSSFind algorithm to its full-bodied counterpart, iRF, to ensure that conclusions drawn from the former extend to the latter.  To concretize these goals (guideline~\ref{g2}), the authors describe the theoretical results they hoped to recover, modeling assumptions and potential violations, and their strategy for comparing iRF to LSSFind.  Because the authors sought to understand the theoretical properties of LSSFind and iRF, their connections to ``reality'' comprised realistic model assumptions and violations; they acknowledge several limitations of their study, including limited tuning of iRF parameters and omission of datasets with multiple assumptions violated (guideline~\ref{g3}).  Theoretical tractability required working with a simplified version of iRF, but as indicated by objective (c), the authors sought to understand the ramifications of this simplification by comparing the test algorithm, LSSFind, against iRF itself as the published state-of-the-art alternative (guideline~\ref{g4}). 

\item[B.] \textbf{List of Ingredients (Data)}: \textit{What should go into it?} \\ The authors generated data using Monte Carlo sampling from an LSS model; thus, the ingredients for their study comprised the LSS models and parameters.  They generated various datasets, including several that violated their modeling assumptions and might have proved inhospitable to the study methods (guideline~\ref{g7}).  The authors also generated input features from real-world data,\footnote{ChIP-seq data as employed for the iRF project (Section~\ref{sec:casestudy}).} producing response values that satisfy some (but not all) conditions of the LSS model (guideline~\ref{g6}).  These data-inspired simulations ultimately did not see publication due to page restrictions but, for transparency, appear in the project GitHub repository.  Because LSSFind acted as a proxy for understanding iRF rather than an algorithm for use by practitioners, the authors considered its performance on real-world data to be of relatively low importance; in this way, LSSFind was indeed deployed in its intended context (guideline~\ref{g5}).  Nonetheless, our understanding of iRF might have benefited from deeper investigation of either algorithm in the intended setting of real-world \textit{omics} data.

\item[C.] \textbf{Preparation Steps (Methods)}: \textit{What steps are required?} \\ The primary methods here included LSSFind itself and its more-complex analogue, iRF (the state-of-the-art competitor, per guideline~\ref{g9}\footnote{Guideline~\ref{g9} recommends comparing against a selection of methods, but the problem statement takes precedence.  The authors choose a single reference method in light of this study's relatively narrow stated purpose and the general lack of existing comparable methods.}).  Both methods depended on tuning parameters, such as the number of iterations and number of bootstrap replicates for iRF and the parameters $\epsilon$ and $\mu$ for LSSFind (see paper for details). The authors implemented LSSFind in Python and used a pre-existing implementation of iRF in the same language.  They described their modeling procedures, including computational specifications and parameters, in the full paper (guideline~\ref{g8}).  The simulation code, encapsulated in Jupyter notebooks, appears online, but would have benefited from more-thorough documentation.  The authors considered two performance metrics suited to both algorithms (guideline~\ref{g10}), including one metric to inform the correctness of their theoretical results and another of more-practical relevance to compare the performance of LSSFind against that of iRF.

\item[D.] \textbf{Staging (Reporting)}: \textit{How will the finished product look if done right?} \\ \cite{LSSFind} reported their results primarily as textual descriptions.  Figures included a graphical illustration of the tree-based algorithm procedure and, in the Supporting Information, bar charts of performance metrics for LSSFind alone or as compared to iRF.  These bar charts are straightforward (guideline~\ref{g11}), but less narrative text and more (varied) figures might have aided readers in digesting the (importance of the) theoretical results (guideline~\ref{g12}). For reproducibility, the authors published all simulation code, including the implementation of LSSFind, on GitHub.  Although they might have documented their code/repository more completely, all evidence suggests that the authors tried to represent their results and conclusions to their audience in a reasonable and honest manner (guideline~\ref{g13}).
\end{enumerate}

\subsection*{\texorpdfstring{$\mathcal{MERITS}$}{MERITS} and Shortfalls}

\begin{itemize}
\item $\mathcal{M}$\textit{odular}: Each simulation occupies a separate Jupyter notebook.  Code shared across simulations, including the implementation of LSSFind, is loaded into each notebook from a joint Python script to limit code redundancies.  This implementation could have benefited from a stand-alone package for public dissemination (but note that the project goals were primary theoretical rather than method development).

\item $\mathcal{E}$\textit{fficient}: The authors of LSSFind sought to create examples complex enough to inform their theoretical investigation of tree-based interaction recovery while minimizing code complexity and run-times.  Notably, the iRF procedure is computationally more efficient, but theoretically less tractable, than LSSFind.  The authors tried to improve the efficiency of LSSFind by searching over a restricted subset of candidate interactions, but could have invested more effort in optimizing the algorithm if method development had been a more-central goal of the project.

\item $\mathcal{R}$\textit{ealistic}: Due to the theoretical nature of the problem, generating context-appropriate data was unusually straightforward.  The authors generated datasets that violated each of their three modelling assumptions, but did \textit{not} consider violating combinations/subsets of assumptions (a trade-off with computational Efficiency). The authors also performed an analysis using real-world ChIP-seq data to assess the performance of their methods in the intended deployment context.\footnote{These results appear in the GitHub repository but not the journal publication.} 

\item $\mathcal{I}$\textit{ntuitive}: The LSSFind authors designed their simulations to be as straightforward as possible, limiting the number of settings explored to those that would provide evidence to inform their theoretical investigation without overwhelming the audience.  This strategy resulted in a more-streamlined, but less extensive, simulation: The authors could have considered more combinations of parameters or model assumptions, at the possible cost of lower intelligibility.  The authors could have communicated more clearly the results arising from real-world data, which, in retrospect, even in the online repository were less detailed and clear than we would recommend.

\item $\mathcal{T}$\textit{ransparent}: The authors published all simulation code and data publicly on GitHub, including supplemental project materials that ultimately did not fit the page restrictions of the parent journal.  However, in retrospect, the authors could have provided more documentation of their code.  They also did not publish the study materials for facile public consumption in an {\tt R} package or Python library, but only as Python scripts and notebooks within the GitHub repository.

\item $\mathcal{S}$\textit{table}: The simulations in this project were straightforward by design, and thus the authors encountered relatively few problems with respect to stability across results.  That said, they could have demonstrated the stability of results in more detail: for example, in terms of the methods' tuning parameters (particularly $\epsilon$ and $\mu$, for which they performed no stability analysis), or by addressing any effect of Monte Carlo repetitions (40 in this case).  These considerations would be crucial to a data scientist using LSSFind in practice.  However, because the authors investigated LSSFind primarily as a test-bed for better understanding the more-complicated iRF procedure, and because the investigation was primarily theoretical in nature, they chose to weight the desideratum of Intuitiveness higher than that of Stability.
\end{itemize}

\section{Appendix: Voting-Rights Lawsuit (applied simulation case study)}\label{sec:VRL}

Kentucky's U.S. Senate primary election was scheduled for May 2020, just two months after the World Health Organization deemed COVID-19 a ``pandemic'' and the U.S. government declared a national emergency.  To shield elderly poll workers from exposure to the virus, election officials suspended Kentucky's traditional practice of voting in person at local precinct stations. Instead, Kentucky offered a battery of early and absentee voting options, while reducing in-person Election Day voting to a single, centralized county location.  

\medskip Litigants argued that the reduction in voting locations would disenfranchise citizens -- including Black voters supportive of a candidate seemingly poised to defeat the Democratic Party's favored nominee~\citep{Voss24}. As part of that litigation, \cite{votingRights} sought to anticipate the effects of these unprecedented measures on voter turnout. Herein we focus on the authors' use of simulation to estimate the depressive effect on political participation of relying on county-wide vote centers.  We dissect this simulation through the lens of our framework, demonstrating the PCS primer's utility for a non-methodological, Social Science study.  As per guideline (\ref{g1}), the authors designed their simulation upfront, with no major modifications over the course of the study. 

\subsection*{Recipe Components}

\begin{enumerate}
\item[A.] \textbf{Recipe Header (Problem Statement)}: \textit{What are we trying to produce?} \\ The simulation by \cite{votingRights} sought to anticipate two outcomes: (1) How much would using vote centers suppress political participation? and (2) Would such voter demobilization differ by racial group? (guideline~\ref{g2}).  Traditionally, voters live within 1 mile from their voting location, whereas requiring constituents to travel further to reach a county polling station might suppress voter turnout.  Existing literature documents that distance from polling stations poses a significant and possibly prohibitive obstacle to many voters, one that \cite{Gimpel03} show to differ by type of community (guideline~\ref{g4}).  The real-world implications of this effect of travel distance are fairly evident, as policies that erode the franchise for African-American voters could alter election outcomes and violate constitutional law (guideline~\ref{g3}). 

\item[B.] \textbf{List of Ingredients (Data)}: \textit{What should go into it?} \\ \cite{Haspel05} offer a model for turnout as a function of distance from polling location.  Using this model's estimates, derived from individual-level Atlanta voting data, a simulation can contrast travel distances mandated by use of a centralized polling station against those required under alternative policies, thereby forecasting the effect of reducing polling stations (guideline~\ref{g5}).  Available data include the locations of all pre-pandemic voting locations, as well as the approximate locations of voting-age citizens (captured using the geographic centroids for their Census block groups, a unit for which the Census Bureau provides racial demographics) (guideline~\ref{g6}). Indirectly, therefore, the pairwise distances between potential polling places and constituent residential locations also are known.  Rather than contrast the vote-center approach to one using all historical voting locations -- prohibitive, in light of pandemic-era staffing scarcity -- a more-realistic alternative would have been to open ten or at least five county voting locations. Still, also calculating pre-pandemic travel distances affords a full estimate of the new policy's impact (guideline~\ref{g7}).

\item[C.] \textbf{Preparation Steps (Methods)}: \textit{What steps are required?} \\ The investigators did not know which precinct stations could have been left open. Not all locations would have supported the large-scale voter traffic incurred by operating fewer stations, and some refused to host the election for fear of the virus.  Thus, the simulation entailed randomly selecting precinct locations to open\footnote{The planned, centralized county station plus four or nine additional locations (depending on the policy).} and estimating the distribution of resulting distances that voters would have needed to travel under each simulated batch of polling stations (guideline~\ref{g8}).  The authors compared the average minimum travel distance across policies, as well as the distributions of distances (guideline~\ref{g10}).  The resultant effect on voter turnout could be estimated by applying the state-of-the-art \cite{Haspel05} model to those distances (guideline~\ref{g9}).

\item[D.] \textbf{Staging (Reporting)}: \textit{How will the finished product look if done right?} \\ \cite{votingRights} performed independent, identical simulations for seven of Kentucky's most-populous counties.  For each policy and county, the authors provide a consistent set of summaries, including frequency tables for Census block-groups binned by average minimum travel distance, and histograms for the sampling distributions of minimum travel distances (guideline~\ref{g11}). To gauge disparate racial effects, they provided similar displays stratified by demographic group (guideline~\ref{g12}).  The authors used the best known model for disenfranchisement as a function of distance, but acknowledge frankly a few potential improvements to their study\footnote{Two extensions prevented by the litigation's quick deadline were estimating Manhattan rather than Euclidean distances (to emulate driving patterns) and soliciting guidance on viable polling stations.} (guideline~\ref{g13}).
\end{enumerate}

\subsection*{\texorpdfstring{$\mathcal{MERITS}$}{MERITS} and Shortfalls}

\begin{itemize}
\item $\mathcal{M}$\textit{odular}: \cite{votingRights} compartmentalize their code into separate files, one per analysis step: data cleaning and reformatting; simulation; and visualization.  The study comprises one simulation per county, with identical procedures.  Additional counties could be added with minimal difficulty if analogous data became available.

\item $\mathcal{E}$\textit{fficient}: The simulation design is efficient in both its conception and implementation.  The code is concise and can be executed to completion in a matter of minutes.  

\item $\mathcal{R}$\textit{ealistic}: \cite{votingRights} incorporated real-world data (Census block-group demographics and pre-pandemic precinct stations) and best established theory (model by \citealt{Haspel05}).  They articulate potential areas for improvement, as by incorporating expert knowledge, if such existed, to sample only among stations realistically equipped for the increase in traffic necessitated by consolidating in-person voting to fewer locations.  They also propose improvements to simulation realism by measuring distances using the central mass of a block-group rather than its geographic center, and the Manhattan rather than Euclidean distance metric to emulate travel patterns.

\item $\mathcal{I}$\textit{ntuitive}: The authors chose a simple simulation design for ease of communication to a lay audience.  Presentations (tables and histograms) are similarly straightforward and displayed in close proximity to support comparisons across alternative policies.

\item $\mathcal{T}$\textit{ransparent}: The authors describe their study fully and clearly in the published report, with justification of all study design decisions.  They prespecified all design elements to avoid biasing results, with no deviations required during implementation.

\item $\mathcal{S}$\textit{table}: \cite{votingRights} set random seeds to support the reproducibility of their results, and executed the study under multiple seeds (results not shown) to mitigate the role of luck in their conclusions.  Using multiple simulated replicates allows for a robust characterization of the sampling distributions.  However, the best model in use was developed in reference to Atlanta, GA, and thus may not represent Kentucky perfectly; future extensions of the study might include a stability analysis against model mis-specification as well as across the above improvements to study realism. 
\end{itemize}

\section{Appendix: Further Reading}\label{sec:further}

For readers designing empirical experiments to accompany simulation studies, we recommend a recent contribution in the Social Sciences~\citep{Blair} that, despite tackling a different manner of study design, resonates strongly with our Data Science philosophy.  The authors set out two conceptual and two pragmatic components: The former, Models (of how the world might act) and Inquiries (research questions, stated in terms of the models), are key elements of our problem statement; the latter comprise Data strategies (procedures for gathering information about the world) and Answer strategies (similar to our Methods).  Their accompanying design principles recommend planning studies deliberately upfront (while being responsive to intermediate results) and minding one's audience.

\medskip Other research groups have leveraged their expertise in specific types/aspects of simulation to develop specialized guidance documents or checklists; for readers with related research agendas, this more-granular form of counsel might complement our primer. 

\medskip \cite{Monks} reviewed, and found lacking, a literature of reporting paradigms that ran the gambit from too broad (MIASE, per \citealt{MIASE}) or outmoded (GASS, per \citealt{GASS}) to too specific (MMRR, per \citealt{MMRR}; ODD, per \citealt{ODD1, ODD2}; CHEERS, per \citealt{Cheers}) or ill-suited to simulation (GLP4OPT, per \citealt{GLP}).  Their trio of STRESS checklists sought to improve reporting in Operations Research of agent-based, discrete-event, and system-dynamics simulations, respectively.  Principles common across the checklists include articulation of study purpose and intended usage (problem statement); rationale and implementation details for both data and methods; and documentation for reproducibility (reporting).

\medskip The Minimal Information about CLinical Artificial Intelligence Modelling (MI-CLAIM) checklist, like our primer, draws some inspiration from the PCS framework~\citep{MI-CLAIM}.  It enumerates specific elements of a simulation that the investigator ought to report, spanning five pillars that resemble our simulation ``recipe'' with an expanded methods component: high-level study design (problem statement); choice or generation of data (data); models, model performance, and model examination (methods); and efforts for reproducibility (reporting).  MI-CLAIM is best suited to studies with a clinical bent.

\medskip For readers seeking tangible/technical advice for writing modular and efficient code, \cite{factor12} provides a set of twelve highly granular recommendations for writing service-oriented code.  Likewise, \cite{PragmatProg} and \cite{PragmatML} provide slightly higher-level advice on building code and machine-learning pipelines that are spiritually very consistent with our framework.  Taking a further step back, \cite{RealWork} advises thoughtfully on the process and practicalities of translating between real-world problems and data/coding solutions.

\medskip Finally, we acknowledge two computational endeavors that share our analogy if not precisely our ambitions.  The {\tt R} package {\tt recipes}, under the {\tt Tidymodels} umbrella, provides tools for preprocessing one's data prior to fitting a model~\citep{tidyRecipes}.  Here the eponymous recipe refers to a sequence of preprocessing steps (\textit{e.g.}, transforming or binning predictor variables).  This package might aid readers seeking to implement our primer with clean and reliable code.  Meanwhile, the book series \textit{Numerical Recipes} treats a range of topics related to algorithms and numerical analysis, describing the fundamentals of the techniques and providing implementations in coding languages that vary by publishing edition~\citep{numericalRecipes}.  The title connotes a ``cookbook'' for numerical computation, allowing readers to delve into each algorithm's ingredients and preparation rather than encountering a mere ``menu'' of finished products.  The authors share some of our guiding philosophy, seeking to promote a deeper understanding of methods and to demonstrate that computational methods can be efficient and intuitive rather than opaque black boxes.

\end{document}